\documentclass[conference]{IEEEtran}
\IEEEoverridecommandlockouts
\usepackage{cite}
\usepackage{amsmath,amssymb,amsfonts}
\usepackage{xspace}
\usepackage{subfiles}
\usepackage{booktabs}
\usepackage{subfiles}
\usepackage{graphicx}
\usepackage{multicol}
\usepackage{subcaption}
\usepackage{array}
\usepackage[wby]{callouts}
\usepackage{url}
\usepackage{mathtools}
\usepackage{chngcntr}
\usepackage{algorithm}
\usepackage{titlesec}
\usepackage{framed}
\usepackage{tcolorbox}
\usepackage{adjustbox}
\usepackage{hyperref} 
\usepackage{array, booktabs, makecell}
\usepackage{siunitx, mhchem}
\usepackage{csquotes}
\usepackage{graphicx}
\usepackage{textcomp}
\usepackage{xcolor}
\usepackage{bm,amsmath}
\usepackage{amsfonts} 
\usepackage{physics}
\usepackage{algorithm}
 \usepackage{algpseudocode}
 \usepackage[a4paper,
            bindingoffset=0.2in,
            left=1.7cm,
            right=1.7cm,
            top=1.9cm,
            bottom=2.7cm]{geometry}
 \usepackage{algorithmicx}
 \algdef{SE}[DOWHILE]{Do}{doWhile}{\algorithmicdo}[1]{\algorithmicwhile\ #1}%

\algrenewcommand\algorithmicrequire{\textbf{Input:}}
\algrenewcommand\algorithmicensure{\textbf{Output:}}
\makeatletter
\def\ps@IEEEtitlepagestyle{%
\def\@evenfoot{}%
}
\def\BibTeX{{\rm B\kern-.05em{\sc i\kern-.025em b}\kern-.08em
    T\kern-.1667em\lower.7ex\hbox{E}\kern-.125emX}}
\begin{document}

\title{Ensemble Method for System Failure Detection Using Large-Scale Telemetry Data
}
\author{\IEEEauthorblockN{Priyanka Mudgal}
\IEEEauthorblockA{
\textit{Intel Corporation, USA}\\
priyanka.mudgal@intel.com}
\and
\IEEEauthorblockN{Rita Wouhaybi}
\IEEEauthorblockA{
\textit{Prev. Intel Corporation, USA}}
}

\maketitle

\begin{abstract}
The growing reliance on computer systems, particularly personal computers (PCs), necessitates heightened reliability to uphold user satisfaction. This research paper presents an in-depth analysis of extensive system telemetry data, proposing an ensemble methodology for detecting system failures. Our approach entails scrutinizing various parameters of system metrics, encompassing CPU utilization, memory utilization, disk activity, CPU temperature, and pertinent system metadata such as system age, usage patterns, core count, and processor type. The proposed ensemble technique integrates a diverse set of algorithms, including Long Short-Term Memory (LSTM) networks, isolation forests, one-class support vector machines (OCSVM), and local outlier factors (LOF), to effectively discern system failures. Specifically, the LSTM network with other machine learning techniques is trained on Intel® Computing Improvement Program (ICIP) telemetry software data to distinguish between normal and failed system patterns. Experimental evaluations demonstrate the remarkable efficacy of our models, achieving a notable detection rate in identifying system failures. Our research contributes to advancing the field of system reliability and offers practical insights for enhancing user experience in computing environments.
\end{abstract}


\section{Introduction}
System failures primarily stem from system errors, with each critical system error having the potential to trigger system shutdowns or reboots in an attempt to rectify the issues \cite{ref2}. These errors can result in substantial financial losses and significant harm to critical information technology (IT) infrastructure. For instance, in the context of Windows, the commonly encountered "blue screen of death" (BSOD) often appears, disrupting device usage \cite{ref2}. Meanwhile, other non-critical errors may arise without causing any disruption to the device. The underlying causes of these errors may originate from various sources, including system hardware or software issues, and can differ based on factors such as system metrics, hardware configurations, and environmental conditions like temperature or frequency of usage, all of which directly contribute to system failures.

Hence, it's crucial to keep a close eye on the health of personal computers (PCs). This monitoring protocol should seamlessly integrate into the product development lifecycle and persist even after PCs have been distributed to end users worldwide. Whether during development or post-deployment, monitoring hinges on telemetry, a method of gathering and retaining data from remote systems. Especially advantageous for clusters or extensive arrays of systems within IT infrastructure, monitoring streamlines the collection of telemetry data for a range of objectives, such as identifying system errors, predicting potential issues, and overseeing overall system health.

This paper utilizes telemetry data from client PCs, focusing on CPU metrics \cite{708428, ref1, 485891, Hastie2009, CHAKRABORTY2011561, Jin2010, 4781136, Hady2013}. The key metrics include CPU utilization, memory utilization, disk utilization, CPU temperature, process type, number of cores, age of the system, and active applications. There are three techniques for training error detection models: supervised, unsupervised, and semi-supervised \cite{708428, ref1, 485891, Hastie2009, CHAKRABORTY2011561, Jin2010, 4781136, Hady2013}. Supervised learning uses labeled data for both normal and anomalous events but requires balanced datasets \cite{485891}. Unsupervised methods identify patterns and flag deviations as anomalies \cite{Hastie2009}. Semi-supervised learning involves training with a small amount of anomalous data \cite{Hady2013}. These techniques find practical application in various domains, including PC error detection using telemetry and industrial settings \cite{708428, ref1, Hastie2009, CHAKRABORTY2011561}.

In this study, we employ an unsupervised learning approach and introduce three distinct ensemble models utilizing long short-term memory (LSTM) \cite{10.1162/neco.1997.9.8.1735}, along with either isolation forest \cite{4781136}, one-class support vector machine (OCSVM) \cite{1437839}, or local outlier factor (LOF) \cite{10.1145/342009.335388}. Our findings indicate that LSTM paired with isolation forest surpasses the performance of LSTM with OCSVM and demonstrates comparable performance to LSTM combined with LOF.

\begin{figure}
\centering
 \vspace{-0.8cm}
\includegraphics[width=1\columnwidth]{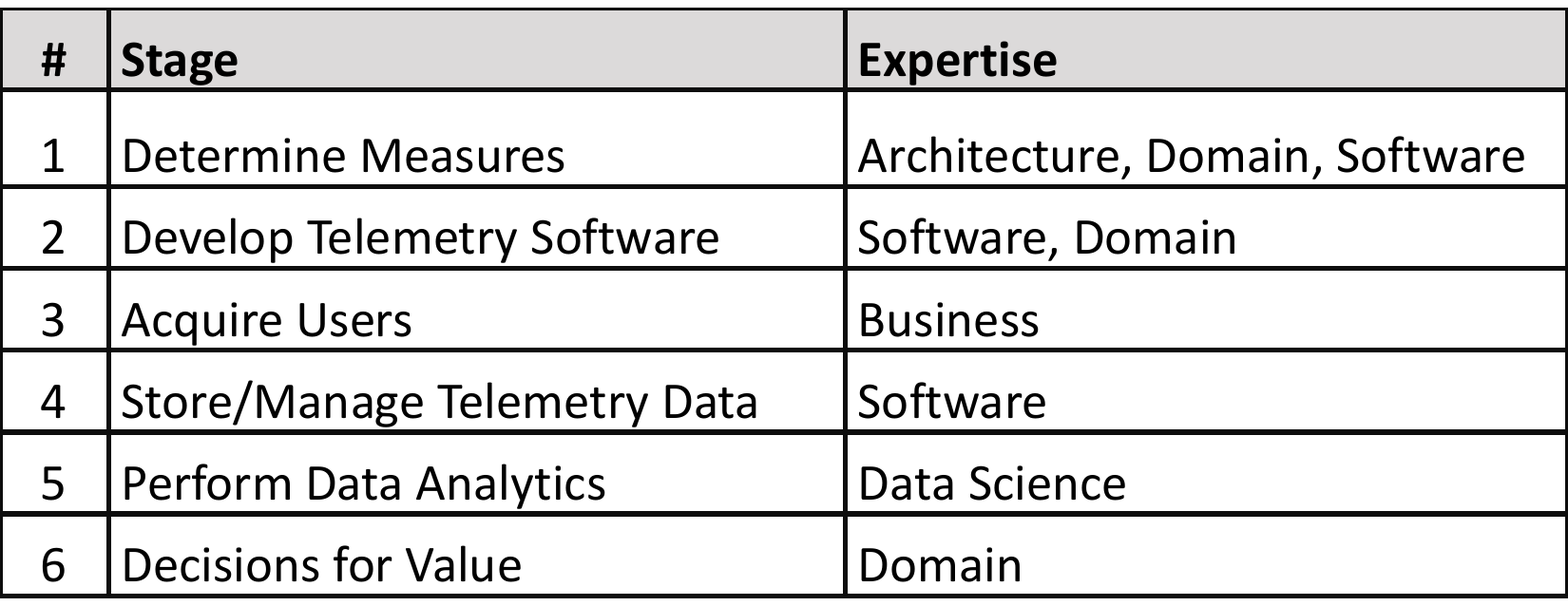}
 \vspace{-2.5cm}
\caption{\label{fig:telemetrystages}Platform Stack and examples of measures}
\end{figure}

\section{\textbf{Background}}

\subsection{\textbf{Telemetry Framework}}

To extract value from telemetry, a comprehensive framework has been outlined by Kwasnick et al. \cite{8993579}, where the authors describe telemetry as a framework comprising six distinct stages, illustrated in Fig. \ref{fig:telemetrystages}. Each stage demands domain expertise to effectively recognize, collect, and refine telemetry data for its subsequent application in data analytics, thereby facilitating informed decision-making tailored to specific domains. Subsequently, we explore the nuances associated with each stage within the framework of product health monitoring.

Stage 1 involves selecting crucial telemetry metrics, considering factors like user privacy and data size limits \cite{8993579, 7936346}. Stage 2 entails the precise gathering and uploading of data by the telemetry collector, adhering to privacy and security standards \cite{8993579, 7936346}. Stage 3 focuses on acquiring user permission for telemetry software download \cite{8993579, 7936346}. Stage 4 addresses data management methods, particularly for managing vast amounts of data with cloud solutions \cite{8993579, 7936346}. Stage 5 discusses data analytics techniques, including correlation analysis and machine learning methods like we use ensemble LSTM \cite{10255554, 10.1162/neco.1997.9.8.1735} in our work. Finally, Stage 6 emphasizes value extraction from telemetry data through informed decision-making and gaining insights into product error behavior for future development \cite{10255554}.

\section{\textbf{Method}}

In our work, we outline our experiments directed towards identifying system errors occurring on end-user systems. Initially, we collect metrics accessible through the Intel® Computing Improvement Program (ICIP) telemetry software \cite{7936346}. ICIP serves as a telemetry software tool for monitoring product health, provided to users upon visiting www.intel.com for driver downloads. Meeting the standards for telemetry software, ICIP ensures privacy, security, and minimal resource consumption. Subsequently, we analyze these metrics to compile the dataset, integrating the specified metrics with system errors. Finally, we train our network using the preprocessed dataset to detect system errors. A comprehensive description of the entire process is provided in the subsequent sections.

\subsection{\textbf{Data Collection}}

The dataset consists of information sourced from systems that have willingly opted to participate in data collection and analytics (DCA) via ICIP. Principally, this data encompasses client PCs featuring various generations of Central Processing Units (CPUs). DCA acquires data from machines exclusively during their operational phases, referred to as the S0 state, and collects it at regular intervals, typically every 5 seconds. On-device aggregation of data occurs every 24 hours, with the aggregated data being uploaded to the datastore when the system is active and connected to the network. Data is only accessible for the days when the machine is active, specifically in the S0 state, for at least a few seconds. The dataset incorporates details regarding machine configurations, memory usage, disk usage, CPU usage, and CPU temperature. We incorporate data spanning 30 days from March 2023 to April 2023, utilizing data from systems with at least 10 days of data available, with at least 10\% of the data indicating system errors.

\begin{figure*}
\setkeys{Gin}{width=.25\linewidth} 
\begin{minipage}[t]{2\columnwidth}
  \subfloat[]{\label{fig:pop}\includegraphics{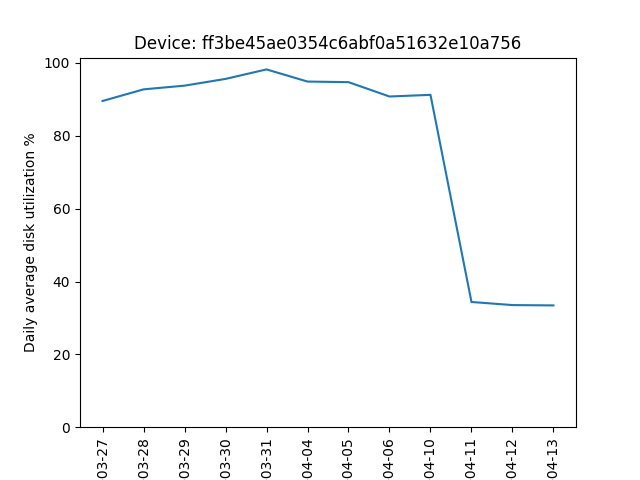}} \hfill
  \subfloat[]{\label{fig:pqua}\includegraphics{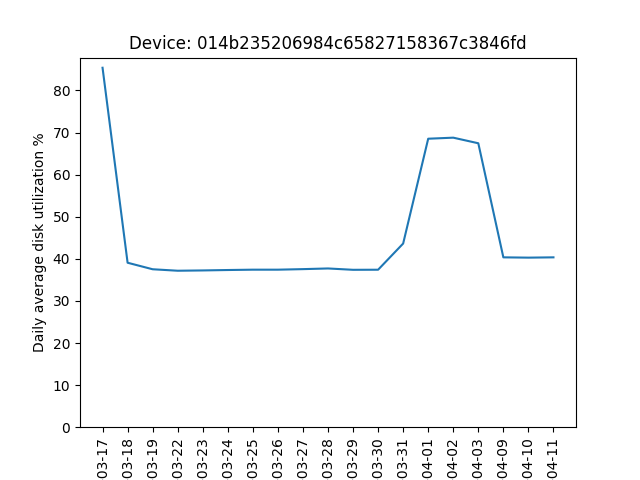}}\hfill
  \subfloat[]{\label{fig:nqua}\includegraphics{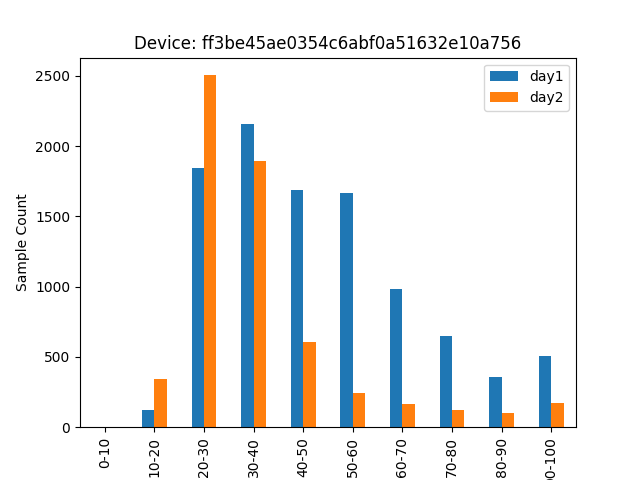}}\hfill
  \subfloat[]{\label{fig:naes}\includegraphics{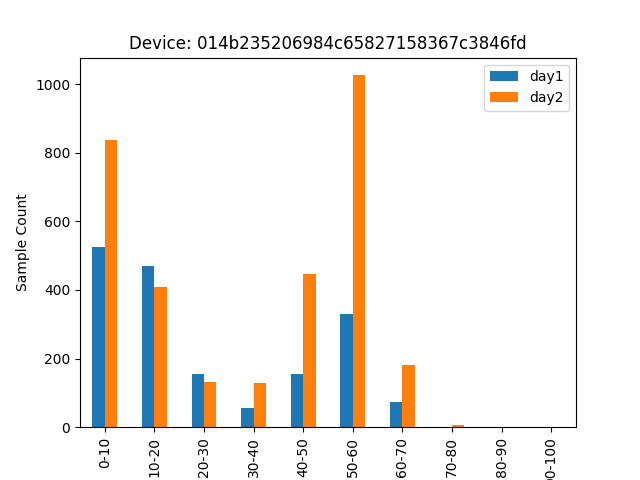}}\hfill
  \caption{Disk utilization (a and b) and CPU utilization (c and d) examples from the ICIP \cite{7936346} dataset} \label{fig:utils}
\end{minipage}\hfill 

\end{figure*}

In our study, we processed the original data, consisting of over 96 columns and approximately 1 million rows \cite{msmetrics}. Daily aggregate metrics including core temperature, power consumption, CPU and memory utilization, disk usage, system age, persona, core count, and processor type are analyzed \cite{msmetrics}. Preprocessing details are provided in Section \ref{sssec:datapreprocessing} \cite{msmetrics}. Selected attributes are listed in Table \ref{tab:telemetryMetrics} and described below, while example plots are shown in Fig. \ref{fig:utils} \cite{msmetrics}. These attributes are hypothesized to correlate with system errors \cite{msmetrics}.

\begin{table}[t]
\setlength{\tabcolsep}{4pt}
\caption{System metrics used in our work.}
\begin{center}
 \scriptsize
 \begin{tabular}{llcccc}\toprule

\thead{Metric} & \thead{Collection}   \\
 \midrule  \addlinespace
CPU Temperature & Model specific register  \\ 
CPU Utilization & Kernel and user mode image load/unload  \\
Memory Utilization & Kernel-mode memory manager   \\
Disk Utilization & Disk I/O events \\
System Errors & Windows error event \\
System Information & CPU machine check architecture \\
          
 \bottomrule

\end{tabular}
\label{tab:telemetryMetrics}
\end{center}
\end{table}

\subsubsection{\textbf{CPU Temperature}}

The CPU reads and stores the temperature of each core \cite{cputemp} with digital thermal sensors \cite{intelptm}. ICIP records the temperatures every 5 second when the system is switched on. For our work, we used the weighted average of the daily temperature over all the cores in each PC, for each day reporting.

\subsubsection{\textbf{\textit{CPU Utilization}}}

CPU executes commands, low usage ideal, high for intensive programs. ICIP reads each core utilization \cite{msmetrics} every 5 secs, captures average every 12 hrs.

\subsubsection{\textbf{Memory Utilization}} Memory utilization \cite{memutil} refers to the average usage calculated from the percentage of available memory in use at any given time. ICIP captures the memory utilization in each PC every 12 hours.

\subsubsection{\textbf{Disk Utilization}} Disk usage \cite{diskutil} represents the proportion of the hard disk currently utilized by the computer to execute programs and tasks. High or 100\% disk usage can  occur on any category of device, regardless of their age or condition. In this study, we use the average used disk percentage in each PC for daily reporting.

\subsubsection{\textbf{System Information}} System information contains the data about the device's specifications, namely, CPU type, number of cores, age of device, persona using the device, chasis type, etc. We use these details to feed to our machine learning model that could relate to system errors.

\subsubsection{\textbf{System Errors}}

System errors, identified through CPU Machine Check Architecture, are reported to Windows OS along with relevant data \cite{intelptm}. Some error records may have timestamps from the day after if the PC was not booted until then. Our dataset includes about 150 types of system errors with associated bug check codes \cite{bugcheckcode}. Previous studies explored memory and temperature correlations with errors \cite{10255554, 9236782}. However, our work uses the mentioned telemetry data and ensemble machine learning for error detection \cite{msmetrics}.

\subsection{\textbf{Data Preprocessing}} \label{sssec:datapreprocessing}
We employ an extensive dataset sourced from ICIP \cite{7936346} provided by Intel®. Due to the diverse metrics captured at varying frequencies, we undertake several preprocessing steps. The preprocessing methodology for CPU temperature and utilization is identical and described in Algorithm \ref{alg:cputemp}. Initially, we compute the weighted average CPU temperature for all cores on a daily basis. Subsequently, we execute the same process for CPU utilization data. Both CPU temperature and utilization data are segmented into distinct bins. Samples displaying CPU temperatures ranging from 0 to 10\% are assigned to bin 1, while those from 10 to 20\% are designated to bin 2, and so forth. A similar approach is applied to CPU utilization. We preprocess this data by categorizing the metrics into three groups: low, medium, and high. If the combined number of samples in bins 9 and 10 constitutes 80\% or more of the total samples, the metric is labeled as \enquote{high}. Similarly, if the combined number of samples in bins 5, 6, 7, and 8 amounts to 80\% or more of the total samples, the metric is categorized as \enquote{medium}. Likewise, if the combined number of samples in bins 1, 2, 3, and 4 equals 80\% or more of the total samples, the metric is classified as \enquote{low}. The same methodology is applied to CPU utilization.

\begin{algorithm}

\caption{An algorithm to preprocess disk utilization and CPU temperature and utilization data (T)}\label{alg:cputemp}
$status \gets Uncategorized$

\begin{algorithmic}

\Require $T$, \\
where $T$ = \{s, c, $N[b_k]$\},\\
$s$ - system id, \\
$c$ - core,\\
$N[b_k]$ - number of samples in $k^{th}$ bin, \\
where $k = 1,\cdot\cdot\cdot, total \: number \: of \:bins$
 
\Ensure $t$ = \{s\}\\

\State $W=\dfrac{\sum_{i=1}^{n}{w_iX_i}}{\sum_{i=1}^{n}w_i}$ \Comment{Calculate weighted average for all cores} \\
where, $W$ = weighted average,\\ 
$n$ = number of terms to be averaged,\\
$w_i$ = weights applied to x values,\\
$X_i$ = data values to be averaged
\If{$\sum_{j=1}^{m}W[N[b_j]]$ $>=$ $p(\sum_{k=1}^{t}W[N[b_k])$, \\
where, $j = $ 9 or 10 for CPU temperature and utilization; \\
$j = $ 19, $\cdot \cdot \cdot$, 21 for memory utilization; \\
$p = 0.8$ \\
}
\State $status \gets high$
   
\ElsIf{$\sum_{j=1}^{m}W[N[b_j]]$ $>=$ $p(\sum_{k=1}^{t}W[N[b_k])$,\\
where, $j = $ 5,  $\cdot \cdot \cdot$, 8 for CPU temperature and utilization; \\
$j = $ 11, $\cdot \cdot \cdot$, 18 for memory utilization; \\
$p = 0.8$\\
}
\State $status \gets medium$
\ElsIf{$\sum_{j=1}^{m}W[N[b_j]]$ $>=$ $p(\sum_{k=1}^{t}W[N[b_k])$\\
where, $j = $ 1,  $\cdot \cdot \cdot$, 4 for CPU temperature and utilization;\\
$j = $ 1, $\cdot \cdot \cdot$, 10 for memory utilization; \\
$p = 0.8$\\
}
\State $status \gets low$

\EndIf
\end{algorithmic}
\end{algorithm}

Memory utilization for the systems follows a similar organization, albeit with a few adjustments. The bin size for memory utilization is set at 5, wherein samples displaying memory utilization below 5\% each day are allocated to bin 1. Likewise, samples exhibiting memory utilization above 5\% but below 10\% are placed in bin 2, and so forth. Furthermore, memory utilization incorporates an additional bin to account for samples indicating more than 100\% memory utilization, warranting a separate allocation. Consequently, memory utilization encompasses a total of 21 bins. Subsequently, we preprocess this data as outlined in Algorithm \ref{alg:cputemp}. After computing the weighted average for all cores per day, if the collective number of samples indicating memory utilization of 95\% and above accounts for 80\% of the total samples, the memory utilization is categorized as \enquote{high}. Similarly, if the total number of samples indicating memory utilization of 55\% and above but less than 95\% constitutes 80\% of the total samples, the memory utilization is labeled as \enquote{medium}. Finally, if the total number of samples indicating memory utilization below 55\% constitutes 80\% of the total samples, the memory utilization is categorized as \enquote{low}. Additionally, we consider the total number of samples in each of the three categories.

We aggregate disk, memory, CPU utilization, temperature, and system data daily, ensuring accuracy by imputing time intervals for diverse devices \cite{msmetrics}. Error labels are appended to rows indicating system errors, serving for validation \cite{msmetrics}. The dataset grows to over a million rows, with systems operational for at least 10 days included \cite{msmetrics}. Categorical data is transformed into numerical representations to aid prediction model development \cite{msmetrics}.

\subsection{\textbf{Network Architectures}}
This section outlines the system architecture of our proposed model.

\begin{figure}
\setkeys{Gin}{ width=1\linewidth} 
\begin{minipage}[t]{1\columnwidth}
  \vspace{-1.5cm}
  \centering
  {\label{fig:architecture}\includegraphics{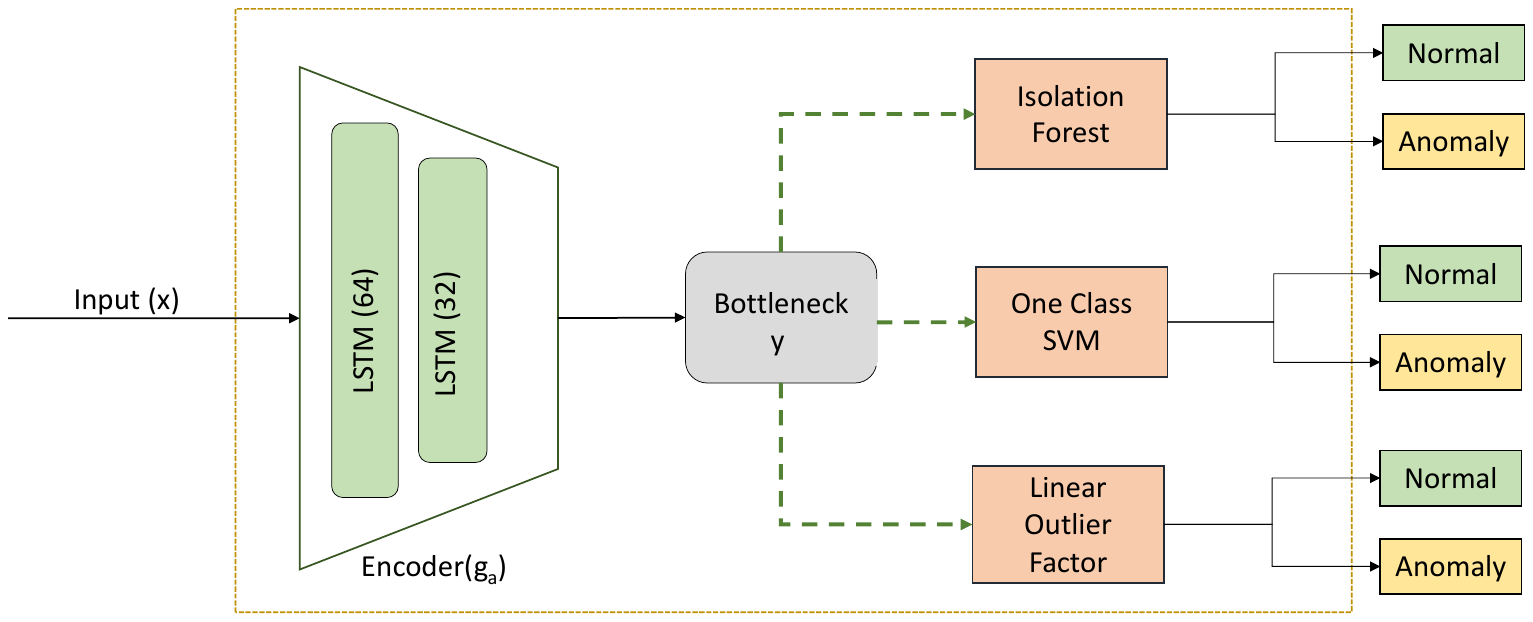}}
 \vspace{-2cm}
   \caption{Proposed ensemble architecture. Overall LSTM is used to encode the telemetry data, which is then fed to three different machine learning models, namely, isolation forest, one-class support vector machine, and local outlier factor.} \label{fig:architecture}
\end{minipage}\hfill 

\end{figure}
\subsubsection{\textbf{Proposed Architecture}}
In this section, we present the proposed architecture designed for identifying anomalies in platform telemetry data. While traditional machine learning techniques may benefit from feature extraction to enhance results \cite{BARANDAS2020100456}, this process typically demands domain expertise.  In contrast, deep learning techniques leverage multi-layer processing to effectively model input features, offering advantages over traditional hand-crafted feature descriptors. In our architecture, we incorporate Long-Short Term Memory (LSTM) \cite{10.1162/neco.1997.9.8.1735} within an autoencoder \cite{bank2021autoencoders} framework, leveraging its proven efficacy in anomaly detection tasks \cite{10.1145/3416013.3426457, bontemps2017collective, 9217754, thi2018oneclass}, and its demonstrated high performance. By using this architecture, the temporal correlation of the platform telemetry data which is often time-series data \cite{Tang2018DeepRN} is leveraged to transform the data in latent space. We use the encoder part of the autoencoder architecture to encode the data in a fixed range feature vector $Y$. The encoder contains two layers of LSTM block. We set the timestamp as one for the LSTM blocks. The final encoded feature $Y$ represent the compressed data. This encoded data $Y$ is subsequently inputted into distinct machine learning models, namely Isolation Forest  \cite{4781136, 8283275, LESOUPLE2021109, iforest_wiki}, One-Class Support Vector Machine (OCSVM) \cite{NIPS1999_8725fb77}, and Local Outlier Factor (LOF) algorithm \cite{10.1145/342009.335388}, individually trained for anomaly classification. These models exclusively utilize normal class data for training, enabling anomalies to be identified as outliers.

Each of these algorithms produce the anomaly score and anomaly. We use these scores to determine if an input data point is classified as an anomaly or normal. These metrics significantly aid in anomaly detection, as the corresponding anomaly score value tends to be notably higher in cases of anomalies.

\subsubsection{\textbf{Experimental Setup}}
We train the LSTM encoder, such that the reconstruction loss is minimum. We use a learning rate of  0.001, batch size of 16, $tanh$ activation function, $huber$ loss function, and Adam optimizer. We train this model for 25 epochs to generate the latent features $Y$. Next, we use the generated features $Y$ to train different machine learning models, namely, isolation forest, OCSVM, and LOF to detect the anomalies in them. We define the contamination value of 0.01 for Isolation Forest and LOF and $nu$ as 0.01 for OCSVM. We train all the models on our system with Intel\textsuperscript{\textregistered} Xeon\textsuperscript{\textregistered} E3-1200 v5 processor and Intel\textsuperscript{\textregistered} Xeon\textsuperscript{\textregistered} E3-1500 v5 processor.
\subsection{\textbf{EVALUATION AND RESULTS}}

This section details the evaluation process of our technique and
shows that our method provides great confidence in classifying the anomalies in telemetry data.

\subsubsection{\textbf{Performance Metrics}}
We evaluate model performance using precision, recall, F1 score, and accuracy. Precision measures true positives over total positive predictions; recall quantifies true positives over actual positives; F1 score balances precision and recall; accuracy measures correct predictions over total instances.

\begin{table}[t]
\setlength{\tabcolsep}{5.5pt}
\caption{Our Results. P represents precision, R represents recall, F1 represents F1-measure, and A represents accuracy.}
\begin{center}
 \scriptsize
 \begin{tabular}{lccccc}\toprule

\thead{Algorithm} & \thead{P $\uparrow$} & \thead{R $\uparrow$} & {F1 $\uparrow$} & \thead{A (\%) $\uparrow$} & \thead{Inference \\ time(s) $\downarrow$ }\\
 \midrule  \addlinespace
LSTM Encoder \\+ LOF & 0.9147 & 0.9389 & 0.9266 & 86.41 & 56.16\\
LSTM Encoder\\ + OCSVM & 0.9149    & 0.9479 & 0.9311 & 87.17 & 93.70\\
LSTM Encoder \\+ Isolation Forest & 0.9149 & 0.9503 & 0.9323 & 87.38 & 3.80\\
          
 \bottomrule

\end{tabular}
\label{tab:scoreSnapshot}
\end{center}
\end{table}

\subsubsection{\textbf{Results and Analysis}}

The results are shown in Table \ref{tab:scoreSnapshot}, where LSTM encoder with Isolation Forest outperforms other techniques in terms of accuracy, precision, recall, and f1-score. Although, these metrics are not far from each other. However, if we consider the inference time, isolation forest is clearly the fastest algorithm. The reported precision, recall, and f1-score are for \enquote{normal} data. These metrics report a very low number for \enquote{abnormalities}. The possible cause may be as the dataset is very imbalanced, where 90\% of the data is \enquote{normal} and 10\% of the data is \enquote{abnormal}. We also trained vanilla machine learning models namely Isolation Forest, OCVSM, and LOF on the same dataset. Although the models yield comparable accuracy, our proposed approach reduced both training and inference time by 1.5$\times$ compared to vanilla machine learning methods.

\section{\textbf{Discussion}}

A hypothesis positing an association between system utilization, temperature, and system errors suggests the potential for system malfunction. The conjecture is that elevated system metrics, particularly those detected by the CPU, may signal a marginality wherein excessive errors could surpass the correction logic's capacity. In this study, we examined all system errors, system utilization, and temperature. However, future investigations could yield further insights by analyzing underlying information from corrected and uncorrected OS and CPU error events \cite{crashlogs, bugcodes}. Additionally, exploring other system metrics such as frequency and OS states during error events, if available, could provide valuable insights. Furthermore, in future research, modern deep learning techniques such as transformers or large language models could be employed, incorporating additional features, including those from the temporal domain. Integrating time-dependent information may offer deeper insights into the underlying causal mechanisms. Our work sets the stage for researchers in this field to explore these possibilities further, potentially leveraging our preprocessing methodology to enhance results.

\section{\textbf{Conclusion}}
In our work, we focus on detecting system errors on end user systems through an ensemble architecture combining deep LSTM and various machine learning techniques. Initially, we collect metrics accessible via Intel® Computing Improvement Program (ICIP) telemetry software \cite{7936346}, a tool for monitoring product health provided to users when they download drivers from www.intel.com. Subsequently, we process these metrics to construct the dataset, integrating specified metrics with system errors. Finally, we train our network using the preprocessed dataset to identify system errors. Our models exhibit remarkable efficacy, achieving a notable detection rate in pinpointing system failures.

\section{\textbf{Acknowledgment}}
We gratefully acknowledge the contributions of our colleagues including Bijan Arbab, Swaathi Sampath Kumar, and Joshua Boelter.

\bibliographystyle{IEEEtran}
\bibliography{ipbib}

\end{document}